\documentclass[english]{article}
\usepackage[T1]{fontenc}
\usepackage[latin1]{inputenc}
\usepackage{amsmath}
\usepackage{graphicx}
\usepackage{amssymb}

\makeatletter

\providecommand{\tabularnewline}{\\}

 \newcommand{\lyxaddress}[1]{
   \par {\raggedright #1 
   \vspace{1.4em}
   \noindent\par}
 }

\usepackage{babel}
\makeatother
\begin{document}

\begin{flushright}
ITP-Budapest 621\\
September 2005\\
\end{flushright}
\smallskip{}

\begin{center}{\LARGE 
Fermion Condensate Model of Electroweak }

{\LARGE
Interactions
}
\bigskip{}

{\large G. Cynolter$^{*}$, E. Lendvai$^{*}$ and G. Pócsik$^{\dagger}$
\\
}
\end{center}

\lyxaddress{\textit{$^{*}$Theoretical Physics Research Group of Hungarian Academy
of Sciences, Eötvös University, Budapest, 1117 Pázmány Péter sétány
1/A, Hungary}\\
\textit{$^{\dagger}$Institute for Theoretical Physics, Eötvös Lorand
University, Budapest, 1117 Pázmány Péter sétány 1/A, Hungary }}

\begin{abstract}
A new dynamical symmetry breaking model of electroweak interactions
is proposed based on interacting fermions. Two fermions of different
$SU_{L}(2)$ representations form a symmetry breaking condensate and
generate the lepton and quark masses. The weak gauge bosons get their
usual standard model masses from a gauge invariant Lagrangian of a
doublet scalar field composed of the new fermion fields. The new fermion
fields become massive by condensation. It is shown that the new charged
fermions are produced at the next linear colliders in large number.
The model is a low energy one which cannot be renormalized perturbatively.
For the parameters of the model unitarity constraints are presented.
\end{abstract}
One of the most important open questions of particle physics is the
origin of elecroweak symmetry breaking. Scalars have not been seen
in experiments emphasizing the importance of investigating models
beyond the standard one {[}1-6{]}. In particular, technicolor models
\cite{dynsym} embody dynamical symmetry breaking but it is not easy
to fulfill electroweak precision tests. The top quark plays a key
role in top condensate models \cite{top}, which predict too large
masses for the top quark and the Higgs boson. Electroweak symmetry
breaking by the condensate of massive matter vector bosons was also
put forward \cite{vcm04}. 

In this note we propose a new low energy effective model based on
four-fermion interaction of new doublet and singlet fermion fields
forming condensates. These lead to massive fermions in the linearized
approximation \cite{njl}. Non-vanishing lepton and quark masses are
coming from gauge invariant couplings of known and new fermions. The
Kobayashi-Maskawa description is unchanged. The usual gauge boson
masses follow from the gauge invariant interaction of a scalar doublet
composed of new fermion fields. The production cross section of the
new fermions is calculated for the next generation of linear electron-positron
colliders. Unitarity constraints for two particle elastic scattering
processess are presented.

We consider the standard model but replace the usual Higgs sector
with new fermions having nonrenormalizable 4-fermion interactions.
The new fermions are a neutral singlet,\[
\Psi_{S},\qquad T=Y=0,\]
and a weak doublet with hypercharge 1,\[
\Psi_{D}=\left(\begin{array}{c}
\Psi_{D}^{+}\\
\Psi_{D}^{0}\end{array}\right),\quad T=\frac{1}{2},\: Y=1.\]
 $\Psi_{D}^{+}\left(\Psi_{D}^{0}\right)$ is a field of positive (zero)
electric charge. The doublet has the same quantum numbers as the standard
Higgs field. The new fermions are assumed to have 4-fermion interactions
as a result of an unknown new physics. Let us assume that in the vacuum
a symmetry breaking condensate is formed\begin{equation}
\left\langle \overline{\Psi}_{S}\Psi_{D}\right\rangle _{0}=\left\langle \left(\begin{array}{c}
\overline{\Psi}_{S}\Psi_{D}^{+}\\
\overline{\Psi}_{S}\Psi_{D}^{0}\end{array}\right)\right\rangle _{0}\neq0.\label{eq:conddoublet}\end{equation}

By $SU_{L}(2)$ transformations we can always transform the upper
component into 0. By $U_{Y}(1)$ transformations of $\Psi_{D}$ the
condensate can be chosen real. $\overline{\Psi}_{S}\Psi_{D}$ resembles
the standard scalar doublet. The condensate (\ref{eq:conddoublet})
breaks properly the weak $SU_{L}(2)\times U_{Y}(1)$ and respects
the electromagnetic $U_{em}(1)$ symmetry. The strength of the condensate
is

\begin{equation}
\left\langle \overline{\Psi}_{S\alpha}\Psi_{D\beta}^{0}\right\rangle _{0}=a_{3}\delta_{\alpha\beta},\quad\left\langle \overline{\Psi}{}_{S\alpha}\Psi_{D\beta}^{+}\right\rangle _{0}=0,\label{eq:mixed cond}\end{equation}
where $a_{3}$ is real.

In what follows we build up the low energy effective theory. We replace
in the standard model Lagrangian the Higgs sector with gauge invariant
kinetic terms for the new fermions and new 4-fermion interactions,
$L_{\Psi}+L_{f}$,\begin{eqnarray}
L_{\Psi} & = & \phantom+i\overline{\Psi}_{D}D_{\mu}\gamma^{\mu}\Psi_{D}+i\overline{\Psi}_{S}\partial_{\mu}\gamma^{\mu}\Psi_{S}+\nonumber \\
 &  & +\lambda_{1}\left(\overline{\Psi}_{D}\Psi_{D}\right)^{2}+\lambda_{2}\left(\overline{\Psi}_{S}\Psi_{S}\right)^{2}+\lambda_{3}\left(\overline{\Psi}_{D}\Psi_{D}\right)\left(\overline{\Psi}_{S}\Psi_{S}\right),\label{eq:4fermion}\\
L_{f} & = & g_{f}\left(\overline{\Psi}_{L}^{f}\Psi_{R}^{f}\right)\left(\overline{\Psi}_{S}\Psi_{D}\right)+g_{f}\left(\overline{\Psi}_{R}^{f}\Psi_{L}^{f}\right)\left(\overline{\Psi}_{D}\Psi_{S}\right),\label{eq:Yukawa}\end{eqnarray}
here \begin{equation}
D_{\mu}=\partial_{\mu}-i\frac{g}{2}\underline{\tau}\,\underline{A}_{\mu}-i\frac{g'}{2}B_{\mu},\label{eq:covariantd}\end{equation}
$\underline{A}_{\mu,}B_{\mu}$ and $g,\; g'$ are the usual weak gauge
boson fields and couplings, respectively.

$L_{f}$ couples $\Psi_{S,D}$ to the traditional left (right) handed
lepton (or quark) doublet $\Psi_{L}^{f}$(singlet $\Psi_{R}^{f}$)
and it can be extended to three families. The dimensionful coupling
constants can be written as $\lambda_{i}=\frac{\tilde{\lambda_{i}}}{M^{2}}$
and $g_{i}=\frac{\tilde{g}_{i}}{M^{2}}$, where the couplings with
tilde are dimensionless and $M$ can be considered as the scale of
the new physics. $M$ is expected to be a few TeV. The masses of the
weak gauge bosons are coming from effective interactions specified
later. $\lambda_{i},\, g_{f}$ are assumed to be positive.

The mixed condensate (\ref{eq:conddoublet}) generates masses for
the standard quarks and leptons and leads to 4-fermion contact interactions.
For example, for the electron and electron neutrino doublet it reads\begin{equation}
L_{f}=g_{e}\left(\overline{\nu}\, e_{R}\overline{\Psi}_{S}\Psi_{D}^{+}+\overline{e}_{L}e{}_{R}\overline{\Psi}_{S}\Psi_{D}^{0}+\overline{e}_{R}\nu\,\overline{\Psi}_{D}^{+}\Psi_{S}+\overline{e}_{R}e_{L}\overline{\Psi}_{D}^{0}\Psi_{S}\right).\label{eq:Lelectron}\end{equation}
 In the linearized approximation $L_{f}$ generates the electron mass
\begin{equation}
m_{e}=-4g_{e}a_{3},\label{eq:melectron}\end{equation}
and $e^{+}e^{-}\Psi_{S}\Psi_{D}$ type interactions in the physical
gauge $\overline{\Psi}_{S}\Psi_{D}^{+}=0$. Down type quark masses
are generated similarly. For up quarks, as usual, couplings to the
charge conjugate $\tilde{\Psi}_{D}=i\tau_{2}\left(\Psi_{D}\right)^{\dagger}=\left(
\left(\Psi_{D}^{0\,}\right)^{\dagger},
-\left(\Psi_{D}^{+}\right)^{\dagger}\right)$ must be introduced. Introducing nondiagonal quark bilinears, the
Kobayashi-Maskawa mechanism emerges. As in the standard model, from
(\ref{eq:melectron}) for two particles $m_{i}/m_{j}=g_{i}/g_{j}$,
but $a_{3}$ is not determined alone by the Fermi coupling constant.

In order to generate masses for the new fermions, the condensates

\begin{eqnarray}
\left\langle \overline{\Psi}_{D\alpha}^{0}\Psi_{D\beta}^{0}\right\rangle  & = & a_{1}\delta_{\alpha\beta},\nonumber \\
\left\langle \overline{\Psi}_{S\alpha}\Psi_{S\beta}\right\rangle _{0} & = & a_{2}\delta_{\alpha\beta},\label{eq:cond diagonal}\end{eqnarray}
are introduced, where $a_{1},\: a_{2}$ are real. Simple models show
that $a_{1},\, a_{2}<0$ may be assumed. Then, in the linearized approximation
one has

\begin{equation}
L_{\psi}\rightarrow L_{\Psi}^{\mathrm{lin}}=-m_{+}\overline{\Psi_{D}^{+}}\Psi_{D}^{+}-m_{1}\overline{\Psi_{D}^{0}}\Psi_{D}^{0}-m_{2}\overline{\Psi}_{S}\Psi_{S}-m_{3}\left(\overline{\Psi^{0}}_{D}\Psi_{S}+\overline{\Psi}_{S}\Psi_{D}^{0}\right),\label{eq:fermion mass}\end{equation}
with\begin{eqnarray}
m_{\pm} & = & -\left(8\lambda_{1}a_{1}+4\lambda_{3}a_{2}\right)=-2a_{1}\lambda_{1}+m_{1},\nonumber \\
m_{1} & = & -\left(6\lambda_{1}a_{1}+4\lambda_{3}a_{2}\right),\label{eq:mtable}\\
m_{2} & = & -\left(6\lambda_{2}a_{2}+4\lambda_{3}a_{1}\right),\nonumber \\
m_{3} & = & \;\;\;\lambda_{3}a_{3}.\nonumber \end{eqnarray}
 Condensate of the charged $\Psi_{D}^{+}$ would only shift $m_{+},\, m_{1},\, m_{2}$
by $-6\lambda_{1},\,-8\lambda_{1},\,4\lambda_{3}$ times the condensate.
For $g_{f}>0:\: a_{3}<0$ and if $\lambda_{j}>0,\, a_{1,2}<0$ it
follows $m_{+},\: m_{1},\: m_{2},\:-m_{3}>0$. For $\lambda_{3}<0$
positivity of masses is possible but would require extra conditions
on $\lambda_{1},\:\lambda_{2}$. To define physical eigenvalues (\ref{eq:fermion mass})
is diagonalized by the unitary transformation \begin{eqnarray}
\Psi_{1} & = & \phantom{-}c\,\Psi_{D}^{0}+s\,\Psi_{S},\nonumber \\
\Psi_{2} & = & -s\,\Psi_{D}^{0}+c\,\Psi_{S},\label{eq:fermion mixing}\end{eqnarray}
where $c=\cos\phi$ and $s=\sin\phi$, $\phi$ is the mixing angle.
The masses of the physical fermions $\Psi_{1},\:\Psi_{2}$ are

\begin{equation}
2M_{1,2}=m_{1}+m_{2}\pm\frac{m_{1}-m_{2}}{\cos2\phi}.\label{eq:mphys}\end{equation}
The mixing angle is given by \begin{equation}
2m_{3}=(m_{1}-m_{2})\tan2\phi.\label{eq:def phi}\end{equation}
Once $m_{1}=M_{1}$, $m_{2}=M_{2}$, the mixing vanishes, $m_{3}=0$
and vice versa.

For $m_{3}<0$ and $M_{1}>M_{2}$, $\left(M_{1}<M_{2}\right)$ we
have $\sin2\phi<0$ $\left(\sin2\phi>0\right)$. It follows that the
physical eigenstates themselves form condensates since\begin{eqnarray}
c^{2}\left\langle \overline{\Psi}_{1\alpha}\Psi_{1\beta}\right\rangle _{0}+s^{2}\left\langle \overline{\Psi}_{2\alpha}\Psi_{2\beta}\right\rangle _{0} & = & a_{1}\delta_{\alpha\beta},\nonumber \\
s^{2}\left\langle \overline{\Psi}_{1\alpha}\Psi_{1\beta}\right\rangle _{0}+c^{2}\left\langle \overline{\Psi}_{2\alpha}\Psi_{2\beta}\right\rangle _{0} & = & a_{2}\delta_{\alpha\beta},\label{eq:condphys}\\
cs\left\langle \overline{\Psi}_{1\alpha}\Psi_{1\beta}\right\rangle _{0}-cs\left\langle \overline{\Psi}_{2\alpha}\Psi_{2\beta}\right\rangle _{0} & = & a_{3}\delta_{\alpha\beta}.\nonumber \end{eqnarray}
There is no mixed condesate as $\Psi_{1},\:\Psi_{2}$ are independent.
For $a_{3}/cs>0$ $\left(M_{1}>M_{2}\right)$ $\Psi_{1}$ forms the
larger condensate. Combining the equations of (\ref{eq:condphys})
one finds

\begin{equation}
a_{3}=\frac{1}{2}\tan2\phi\left(a_{1}-a_{2}\right).\label{eq:a3rel}\end{equation}

For $a_{1}=a_{2}$ one cannot put $a_{3}\neq0$ for $\cos2\phi\neq0$.
As is seen, (\ref{eq:a3rel}) is equivalent to $\left\langle \overline{\Psi}_{1\alpha}\Psi_{2\beta}\right\rangle _{0}=0.$
Comparing (\ref{eq:a3rel}) to (\ref{eq:def phi}) yields \begin{equation}
m_{1}-m_{2}=\lambda_{3}\left(a_{1}-a_{2}\right).\label{eq:m1m2rel}\end{equation}
Substituting $m_{1}-m_{2}$ from (\ref{eq:mtable}) we are lead to
the consistency condition\begin{equation}
a_{1}\left(\lambda_{3}-2\lambda_{1}\right)=a_{2}\left(\lambda_{3}-2\lambda_{2}\right).\label{eq:consistency}\end{equation}
A consequence is that $a_{1}\neq a_{2}$ goes with $\lambda_{1}\neq\lambda_{2}$.

In the physical spectrum there are three new fermions, a charged one
and two neutral ones. They have many self-interactions given by (\ref{eq:4fermion})
and (\ref{eq:fermion mixing}) like $\Psi_{D}^{+}\Psi_{D}^{+}\Psi_{D}^{-}\Psi_{D}^{-}$,
$\Psi_{D}^{+}\Psi_{D}^{-}\left(\Psi_{1}\Psi_{1},\:\Psi_{2}\Psi_{2},\:\Psi_{1}\Psi_{2}\right)$,
$\Psi_{1}^{4}$, $\Psi_{1}^{3}\Psi_{2},\,\Psi_{1}^{2}\Psi_{2}^{2},\,\Psi_{1}\Psi_{2}^{3},\,\Psi_{2}^{4}$.

The masses of the weak gauge bosons arise from the effective interactions
of an auxiliary composite $Y=1$ scalar doublet,\begin{equation}
\Phi=\left(\begin{array}{c}
\Phi^{+}\\
\Phi^{0}\end{array}\right)=\overline{\Psi}_{S}\Psi_{D}.\label{eq:scalardef}\end{equation}
$\Phi$ develops a gauge invariant kinetic term in the low energy
effective description 

\begin{equation}
L_{H}=h\left(D_{\mu}\Phi\right)^{\dagger}\left(D^{\mu}\Phi\right),\label{eq:L phi}\end{equation}
 $D_{\mu}$ is the usual covariant derivative (\ref{eq:covariantd}).

The coupling constant $h$ sets the dimension of $L_{H}$, $[h]=-4$
in mass dimension. We assume $h>0$. (\ref{eq:L phi}) is a nonrenormalizable
Lagrangian leading to the weak gauge bosons masses and some of the
interactions of the new fermions with the standard gauge bosons. 

In the gauge $\Phi^{+}=0$ $L_{H}$ can be written as\begin{eqnarray}
h^{-1}L_{H} & =\phantom+ & \frac{g^{2}}{2}W_{\mu}^{-}W^{+\mu}\Phi^{0\dagger}\Phi^{0}+\frac{g^{2}}{4\cdot\cos^{2}\theta_{W}}Z_{\mu}Z^{\mu}\Phi^{0\dagger}\Phi^{0}+\label{eq:hLH}\\
 &  & +\left[\partial^{\mu}\Phi^{0\dagger}\partial_{\mu}\Phi^{0}-\frac{i}{2}\frac{g}{\cos\theta_{W}}\left(\partial^{\mu}\Phi^{0\dagger}\right)\Phi^{0}Z_{\mu}+\frac{i}{2}\frac{g}{\cos\theta_{W}}\Phi^{0\dagger}Z_{\mu}\left(\partial^{\mu}\Phi^{0}\right)\right]\nonumber \end{eqnarray}
in terms of the usual vector boson fields.

In the linearized approximation of (\ref{eq:hLH}) we put

\begin{equation}
h\,\Phi^{0\dagger}\Phi^{0}\rightarrow h\left\langle \Phi^{0\dagger}\Phi^{0}\right\rangle _{0}=h\left(16a_{3}^{2}-4a_{1}a_{2}\right)=\frac{v^{2}}{2},\label{eq:phi vev}\end{equation}
leading to the standard masses\begin{equation}
m_{W}=\frac{gv}{2},\qquad m_{Z}=\frac{gv}{2\cos\theta_{W}}.\label{eq:standardm}\end{equation}
$v^{2}$ is, as usual, $\left(\sqrt{2}G_{F}\right)^{-1}$; $v=254$GeV
. The tree masses naturally fulfill the important relation $\rho_{\mathrm{tree}}=1$.
(\ref{eq:phi vev}) and (\ref{eq:a3rel}) impose the restriction\begin{equation}
v^{2}=8h\left[\tan2\Phi\left(a_{1}-a_{2}\right)^{2}-a_{1}a_{2}\right]>0.\label{eq:vsquare}\end{equation}

(\ref{eq:hLH}) describes several nonrenormalizable interactions of
the type $Z,\: Z^{2},\: W^{2}$ times $\Psi_{S}^{2},\:\left(\Psi_{D}^{0}\right)^{2}$,
or $\Psi_{S}^{2}\left(\Psi_{D}^{0}\right)^{2}$. Their strengths in
$h$ units are determined by the gauge coupling constants.

A transparent experimental consequence of the model is provided by
the doublet kinetic term in (\ref{eq:4fermion}) whence we get usual
renormalizable fermion pair couplings to gauge bosons 

\begin{eqnarray}
L^{I} & = & \phantom+\overline{\Psi_{D}^{+}}\gamma^{\mu}\Psi_{D}^{+}\left(eA_{\mu}-e\cot2\theta_{W}Z_{\mu}\right)+\frac{g}{2\cos\theta_{W}}\overline{\Psi_{D}^{0}}\gamma^{\mu}\Psi_{D}^{0}Z_{\mu}+\nonumber \\
 &  & +\frac{g}{\sqrt{2}}\left(\overline{\Psi_{D}^{+}}\gamma^{\mu}\Psi_{D}^{0}W_{\mu}^{+}+\overline{\Psi_{D}^{0}}\gamma^{\mu}\Psi_{D}^{+}W_{\mu}^{-}\right).\label{eq:L ffw1}\end{eqnarray}
 To get the physical interactions for the neutral component, the mixing
(\ref{eq:fermion mixing}) must be taken into account, then\begin{eqnarray}
L^{I} & = & \phantom+\overline{\Psi_{D}^{+}}\gamma^{\mu}\Psi_{D}^{+}\left(eA_{\mu}-e\cot2\theta_{W}Z_{\mu}\right)+\nonumber \\
 &  & +\frac{e}{\sin2\theta_{W}}Z_{\mu}\left(c^{2}\overline{\Psi}_{1}\gamma^{\mu}\Psi_{1}+s^{2}\overline{\Psi}_{2}\gamma^{\mu}\Psi_{2}-sc\left(\overline{\Psi}_{1}\gamma^{\mu}\Psi_{2}+\overline{\Psi}_{2}\gamma^{\mu}\Psi_{1}\right)\right)+\nonumber \\
 &  & +\left[\frac{g}{\sqrt{2}}W_{\mu}^{+}\left(c\overline{\Psi_{D}^{+}}\gamma^{\mu}\Psi_{1}-s\overline{\Psi_{D}^{+}}\gamma^{\mu}\Psi_{2}\right)+h.c.\right].\label{eq:Llmix}\end{eqnarray}
The interaction $L_{f}$ in (\ref{eq:Lelectron}) turns out to be
very weak. Indeed, from (\ref{eq:melectron}) and (\ref{eq:phi vev})
we have an upper bound for $g_{e}$, $g_{e}\leq\sqrt{2h}\frac{m_{e}}{v}=\sqrt{2h}g_{e}^{SM}$,
which is suppressed by two factors of the scale of new physics compared
to the standard model value $g_{e}^{SM}$.

To test the model at the forthcoming accelerators we consider the
productions of new fermion pairs in electron-positron annihilation.
It is most useful to investigate the case of a charged new fermion
pair, we denote this $D^{+}D^{-}$.

The contact graph from (\ref{eq:Lelectron}) yields the cross section 

\begin{equation}
\sigma\left(e^{+}e^{-}\rightarrow D^{+}D^{-}\right)=\frac{g_{e}^{2}}{16\pi}s\sqrt{1-4\frac{m_{+}^{2}}{s}}\left(1-\frac{5}{2}\frac{m_{+}^{2}}{s}\right),\label{eq:sigmaee}\end{equation}
where $s$ is the centre of mass energy squared. The cross section
is negligible at moderate $s$. For example at $h\sim\left(2TeV\right)^{-4}$,
$\sqrt{s}=1$TeV it is still at the order of $10^{-13}$fb.

We expect a higher number of events from the photon and Z exchange
processes $e^{+}e^{-}\rightarrow\gamma,Z\;\rightarrow D^{+}D^{-}\!.$
The usual SM coupling at the $e^{+}e^{-}Z$ vertex is\[
i\frac{g}{2\cos\theta_{W}}\gamma_{\mu}\left(g_{V}+\gamma_{5}g_{A}\right),\;\mathrm{where}\quad g_{V}=-\frac{1}{2}+2\sin^{2}\theta_{W},\; g_{A}=-\frac{1}{2}.\]
\begin{figure}
\begin{center}\includegraphics[%
  scale=0.4,
  angle=270]{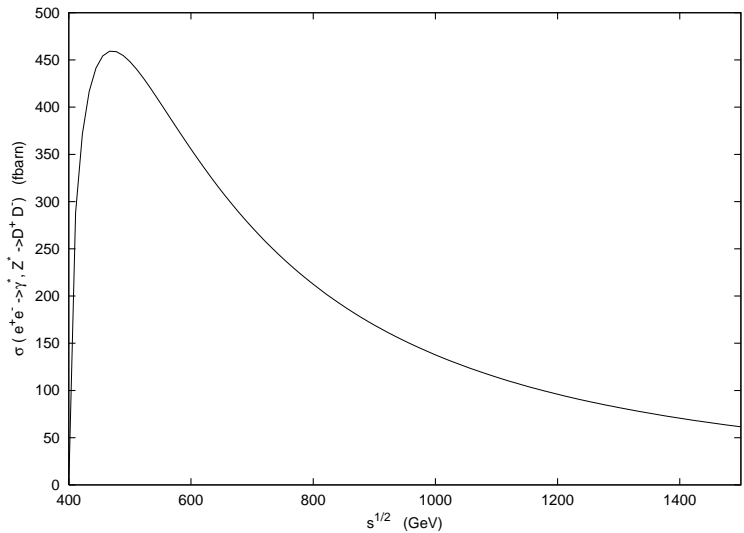}\end{center}

\caption{Cross section of $D^{+}D^{-}$ production at electron-positron collider
vs. $\sqrt{s}$ for $m_{+}=200$GeV}
\end{figure}
\\
By making use of (\ref{eq:Llmix}) one obtains the cross section\begin{eqnarray*}
\sigma\left(e^{+}e^{-}\rightarrow D^{+}D^{-}\right)=\frac{1}{16\pi}\sqrt{1-4\frac{m_{+}^{2}}{s}}\frac{1}{s}\left|M\right|^{2} & ,\end{eqnarray*}
\begin{eqnarray}
\left|M\right|^{2} & = & \phantom+\frac{4}{3}e^{4}\frac{s+2m_{+}^{2}}{s}+\frac{2}{3}\frac{e^{4}}{\sin^{2}\theta_{W}\cos^{2}\theta_{W}}g_{V}\frac{s+2m_{+}^{2}}{s-m_{Z}^{2}}+\nonumber \\
 &  & +\frac{1}{12}\frac{e^{4}}{\sin^{4}\theta_{W}\cos^{4}\theta_{W}}\left(g_{V}^{2}+g_{A}^{2}\right)s\frac{s+2m_{+}^{2}}{\left(s-m_{Z}^{2}\right)^{2}},\label{eq: m2}\end{eqnarray}
where the three terms in $\left|M\right|^{2}$ are coming from photon
exchange , photon-Z interference and pure Z exchange. Similar cross
section belongs to the neutral pair productions, too. The cross section
rises fast after the threshold, at high energies it falls off as $1/s$
reflecting that all the interactions are renormalizable in the process.
The cross section is given in Table 1 for a few masses and plotted
versus $\sqrt{s}$ in Fig. 1. At a linear collider of $\sqrt{s}=500$GeV
(TESLA) and integrated luminosity 50 $fb^{-1}$/year a large number
of events is expected. 

\begin{table}[h]
\begin{center}\begin{tabular}{|l|c|c|c|}
\hline 
$m_{+}$( GeV) &
100&
150&
200\tabularnewline
\hline
$\sigma\left(e^{+}e^{-}\rightarrow D^{+}D^{-}\right)$ (fb)&
560&
535&
450\tabularnewline
\hline
\end{tabular}\end{center}

\caption{Cross section of $D^{+}D^{-}$ production at $\sqrt{s}=$500 GeV}
\end{table}

The cross section at $\sqrt{s}=1500$GeV is an order of magnitude
smaller but with an integrated luminosity of 100 $fb^{-1}$ per annum
a large number of events appears and higher mass range can be searched
for.

\begin{table}[h]
\begin{center}\begin{tabular}{|l|c|c|c|c|}
\hline 
$m_{+}$( GeV) &
100&
200&
400&
700\tabularnewline
\hline
$\sigma\left(e^{+}e^{-}\rightarrow D^{+}D^{-}\right)$ (fb)&
62&
61&
60&
32\tabularnewline
\hline
\end{tabular}\end{center}

\caption{Cross section of $D^{+}D^{-}$ production at $\sqrt{s}=$1500 GeV}
\end{table}

Now, we provide a rough estimate of the model parameters by imposing
partial-wave unitarity, $\left|Re\, a_{0}\right|\leq\frac{1}{2}$,
for the $J=0$ partial-wave amplitudes of $2\rightarrow2$ processes
\cite{unit,unitvcm}. Consider $D^{+}D^{-}$ elastic scattering. At
high energies, $s\gg4m_{+}^{2}$, the contact graph is expected to
give the dominant part of $a_{0}$ leading to $\lambda_{1}s\leq8\pi$.
At a maximum possible energy of $s=(1-25)\, TeV^{2}$, $\lambda_{1}\leq(2.5-0.1)G_{F}.$
In case of no mixing we expect a similar upper bound for $\lambda_{2}$,
too. To obtain an upper bound on $h$ let us consider the process
$\overline{\Psi}_{1}\Psi_{2}\rightarrow ZZ$ with longitudinally polarized
$Z$, and (+ -) helicity for the incoming fermions. The relevant contact
interaction follows from (\ref{eq:L phi})\begin{equation}
L_{H}=-Z_{\mu}Z^{\mu}\overline{\Psi_{D}^{0}}\Psi_{S}\frac{hg^{2}a_{3}}{\cos^{2}\theta_{W}}+h.c.\:,\label{eq:28}\end{equation}
where $16a_{3}^{2}h\gtrsim\frac{v^{2}}{2}$. Choosing $\overline{\Psi}_{1}\Psi_{2}$
at asymptotic $s$ the unitarity imposes $c^{2}\left(s^{3}h\right)^{1/2}\leq16\pi v$.
This shows that $h$ is in general a very weak coupling: at the scale
$s=(1-25)\, TeV^{2}$ $c^{4}h\lesssim2\left(1-5^{-6}\right)G_{F}^{2}$.
We checked that $\bar{t}t\rightarrow\overline{\Psi}_{1}\Psi_{2}$
provides a weaker bound. Fixing $\lambda_{1},\,\lambda_{2}$, larger
$a_{1},\, a_{2}$ give larger masses $m_{+},\, m_{1},\, m_{2}$. For
instance, at $\lambda_{3}=0$, $m_{+}=100\;[300]$GeV needs a condensate
$a_{1}$, $\left|a_{1}\right|\geq\left(\frac{1}{60}-\frac{1}{2.4}\right)G_{F}^{-3/2}$$\quad\left[\left(\frac{1}{20}-\frac{5}{4}\right)G_{F}^{-3/2}\right]$
for $\lambda_{1}\leq\left(2.5-0.1\right)G_{F}$. The lower bound for
$\left|a_{2}\right|$ lies not very far from that of $a_{1}$ for
$m_{1}$ not very far from $m_{2}$. It also follows that the leptonic
coupling constant $g_{e}$ is less than its standard model value,
$g_{e}^{SM}\times{\cal O}\left(G_{F}\right)$, if $c$ is not very
small. Finally, since $8\left|a_{3}\right|\geq2^{1/4}\left(G_{F}h\right)^{-1/2}$,
one gets for $s=(1-25)TeV^{2}$ $\left|a_{3}\right|\gtrsim\frac{1}{\sqrt{2}}(1-125)G_{F}^{-3/2}$
at small mixing.

In conclusion, we proposed a new dynamical symmetry breaking of the
electroweak symmetry based on four-fermion interactions of hypothetical
doublet and singlet fermions. Generating masses for all the standard
and new particles follows by condensation. Condensates are free parameters
of the model. The interactions of new fermions tend to be weak, since
the model is a nonrenormalizable effective low energy one. From experimental
pont of view, $e^{+}e^{-}\rightarrow D^{+}D^{-}$ is advantageous
following from purely gauge interactions and it results in a large
number of events at the next generation of linear colliders.

\end{document}